\documentclass[proof]{WileyASNA-v1}

\articletype{Article Type}%

\received{26 April 2016}
\revised{6 June 2016}
\accepted{6 June 2016}

\begin{document}

\title{Black holes in a bouncing universe}

\author[1]{Iara Pintos*}

\author[1]{Daniela Pérez}

\author[1,2]{Florencia L. Vieyro}

\authormark{Iara Pintos \textsc{et al}}

\address[1]{\orgname{Instituto Argentino de Radioastronom\'ia}, (IAR, CONICET/CIC/UNLP), Camino Gral. Belgrano Km 40, Villa Elisa (1894), Buenos Aires, \country{Argentina}}

\address[2]{\orgdiv{Facultad de Ciencias Astron\'omicas y Geof\'isicas}, UNLP, Paseo del Bosque s/n, La Plata, 1900, Buenos Aires, \country{Argentina}}

\corres{*\email{iara@iar.unlp.edu.ar}}

\abstract{Bouncing cosmologies offer an alternative to the standard $\Lambda$CDM model by avoiding the problem of the initial cosmological singularity by construction. In these models, the universe undergoes a contraction phase that begins in a nearly flat and dilute state, followed by a bounce, after which the universe transitions into the expanding phase described by the $\Lambda$CDM model. During contraction, most large-scale structures are expected to be erased. Black holes, however, as shown by several previous investigations, may persist through the bounce.

The goal of this work is to analyze the evolution of a black hole population throughout the contraction, bounce, and expansion phases. Additionally, we investigate how the presence of black holes influences the properties of the background cosmological fluid. To this end, we develop a cosmological model involving two interacting fluids. Our findings indicate that the cosmological fluid alters its properties in a spacetime with fixed geometry.}




\maketitle

\section{Introduction}\label{sec1}

The idea of a cosmological bounce dates back to the early days of modern cosmology, shortly after the expansion of the universe was discovered \citep{tol31, lem27,lem+97}. The first explicit solutions, however, were obtained in the late 1970s \citep{nov+79,mel+79} . But, it was not until the beginning of this century that nonsingular models began to flourish again. While the standard cosmological model with an inflationary phase effectively explains most of the available data \citep{tri+24,lu+22}, the model is not without problems. One of the most pressing is the issue of the cosmological singularity.

Bouncing models are, by construction, singularity free. The mechanism responsible for generating the bounce may arise either from classical or from quantum effects (for details, see the reviews \citep{Novello2008,bat+14}). In these scenarios, the universe begins in a dilute contracting phase, which evolves smoothly into a bounce and subsequently transitions into an expanding phase.

Several studies have investigated the possibility of black hole formation during the contracting phase \citep{car+11,qui+16,che+17,che+23,bar+25}. Moreover, it has been shown that black holes present at the beginning of contraction can survive the bounce \citep{per+22,cor+22}. Dynamical black hole horizons evolve with cosmic dynamics, allowing a wide range of black holes of different sizes to survive the bounce. Such black holes, once in the expanding branch of the universe, could play a role in structure formation and gravitational wave background generation \citep{per+22}.

As previously mentioned, earlier studies primarily focused on whether individual black holes \citep{per+22,cor+22} or multiple black holes \citep{cli+17,col20} could survive a cosmological bounce. In this work, we assume that black holes do persist through the bounce and investigate the evolution of a black hole population across the contracting, bounce, and expanding phases. Furthermore, we examine how the presence of these black holes influences the properties of the background cosmological fluid. We track the evolution of both components over contracting and expanding phases, each lasting approximately 47,000 years, corresponding to the radiation-dominated epoch.

We are interested in analyzing the changes in the properties of the cosmological fluid that arise with the presence of the black holes. Since we will choose a scale factor, the geometry will be fixed and the bounce will be unavoidable. Therefore, we investigate how the cosmological fluid adapts in this physical scenario. 

The article is organized as follows: in Sec. \ref{sec2}, we present the specific bouncing cosmology adapted in our analysis. In Sec. \ref{sec3} we describe the generalized McVittie metric, which is of vital importance since it tells us how is the coupling of the black hole to the background dynamic. In Sec \ref{sec4} we justify the black hole population as a fluid within the previously introduced background cosmology. The results are presented in Sec. \ref{sec6} and we conclude with final remarks in Sec. \ref{sec7}. We also include the \ref{app1}, where we calculate the parameters of the cosmological model, and \ref{app2}, where we analyze the evolution of the cosmological fluid in absence of black holes. 

\section{Cosmological background}\label{sec2}

Bouncing cosmologies are a possible approach to a non-singular cosmology model. In the framework of General Relativity, the avoidance of the singularity at a classical level requires the violation of the strong energy condition, as we show below.

For an energy-momentum tensor of a perfect fluid in its instantaneous rest frame 
\begin{equation}
    T^\mu_\nu = \text{diag}\left(-c^2\rho, P, P, P\right),
\end{equation}
where $\rho$ and $P$ are the density and pressure of the fluid, respectively, the energy conditions can be expressed as:
\begin{itemize}
    \item Null Energy Condition (NEC): for any null vector $k^\mu$ 
    \begin{equation}
         T_{\mu\nu} k^\mu k^\nu \geq 0 \implies  \rho+P \geq 0. 
    \end{equation}
    \item Weak Energy Condition (WEC): for any time-like vector $v^\mu$
        \begin{equation}
         T_{\mu\nu} v^\mu v^\nu \geq 0 \implies  \rho+P \geq 0~~\text{and}~~\rho\geq0. 
    \end{equation}
    \item Strong Energy Condition (SEC): for any time-like vector $v^\mu$ and being $T = T^\mu_\mu$
     \begin{equation}
        \left( T_{\mu\nu} -\frac{T}{2}\,g_{\mu\nu}\right)v^\mu v^\nu \geq 0 \implies  \rho+P \geq 0~~\text{and}~~\rho+3P\geq0. 
    \end{equation}
\end{itemize}
On the other hand, assuming $k = 0$, the Friedmann equations reduce to:
\begin{equation}
      \rho+P=\frac{c^2}{4\pi\,G }\left[\left(\frac{\dot{a}}{a}\right)^2-\frac{\ddot{a}}{a}\right],
\end{equation}
and 
\begin{equation}
    \rho+3P =-\frac{6c^2}{8\pi\,G }\frac{\ddot{a}}{a}.
\end{equation}
The scale factor $a(t)$ must have certain features in order for a bounce to be defined locally. It is necessary that, at the time of the bounce (conventionally at $t=0$), $\dot{a}_b=0$ and $\Ddot{a}_b>0$. Then, this implies that if there exists a bounce, the NEC is violated. The violation of the NEC, necessarily, implies the violation of the SEC, independently of the geometry of the system. These results follow from the Einstein Equations and the energy conditions, and are independent of any particular equation of state.

It is important to emphasize here that the energy conditions only depend on the scale factor. As we fix the geometry of spacetime, the energy conditions do not change with the components of the universe. In another words, whether there is a population of black holes does not impact on the energy conditions. 

For this work, we have chosen a simple scale factor that describes a realistic bouncing cosmological model in the radiation epoch in terms of the conformal time $\eta$ \citep{celani2017}
\begin{equation}\label{factor_de_escala}
      a(\eta) = a_b\left[1+\left(\frac{\eta}{\eta_b}\right)^2\right]^{1/2}.
\end{equation}
It was derived considering quantum corrections to the Friedmann equations; in other words, the bounce occurs due to quantum effects when the curvature of spacetime becomes very large.  The parameters associated are the following: $a_b = 7.41\times10^{-9}$ is a normalization constant such that $a(\pm\eta_i)=1$, where $\eta_i = 2.96\times10^{12}$ s \footnote{The value for the transition time in cosmic time is $t_i \cong 47\,000$ yr \citep{Ryden_2016}.} is the transition time between the radiation-dominated epoch into the matter-dominated epoch, and $\eta_b = 2.19\times10^4$ s is a time scale associated with the bounce. In the \ref{app1} we provide a detailed calculation of these parameters.

The relation between the conformal time $\eta$ and the cosmic time $t$ is given by:
\begin{equation}\label{cambio_de_coord}
    \frac{{\rm d}\eta}{{\rm d}t} = \frac{1}{a(\eta)}.
\end{equation}
Substituting the scale factor \eqref{factor_de_escala}, we obtain
\begin{equation}\label{tiempo_cosmico}
    t = \frac{a_b}{2}\left[\eta_b~ \text{arcsenh}\left(\frac{\eta}{\eta_b}\right)+\eta\sqrt{\frac{\eta^2}{\eta^2_b}+1}\right].
 \end{equation} 
The cosmic time of the bounce is set at $t= 0$ and, this corresponds to the conformal time $\eta= 0$.

\section{Generalized McVittie spacetime}\label{sec3}

The first exact solution to the Einstein Equations was found by Karl Schwarzschild in 1916 \citep{schwarzschild1916b}. It represents the spacetime outside a spherically symmetric body of mass $M$ with null intrinsic angular momentum and electric charge. Years later, other stationary solutions were found. However, these kind of spacetimes are asymptotically flat and are not useful when one does want to study its evolution in a cosmological context. 

To do so, there are other kind of solutions, such as the generalized McVittie spacetime. The metric in isotropic coordinates $(t, r, \theta, \varphi)$ reads \citep{carrera_2010, faraoni_2007, far15}  
\begin{align}
    \text{d}s^2 &= -c^2\left(\frac{1-\dfrac{G\,m(t)}{2c^2r}}{1+\dfrac{G\,m(t)}{2c^2r}}\right)^2 \text{d}t^2 + \\&+a^2(t)\left(1+\frac{G\,m(t)}{2c^2r}\right)^4\big(\text{d}r^2+r^2\text{d}\theta^2+r^2\sin^2\theta\text{d}\varphi^2\big) \nonumber,
\end{align}
where $m(t)$ is a function that depends on the cosmic time $t$. For large values of $r$, the Friedmann-Lemaître-Robertson-Walker (FLRW) metric with $k=0$ is recovered. 

The source of this geometry is given by an imperfect fluid of the form 
\begin{equation}
    T_{\mu\nu} = \big(\rho+P\big)u_\mu u_\nu+Pg _{\mu\nu} + q_\mu u_\nu + q_\nu u_\mu,
\end{equation}
where $\rho$ is the density, $P$ the pressure, and $q_\mu$ is a purely spatial vector field that represents the current density of heat. It is assumed that the fourth velocity of the fluid $u^\mu$ only has a time component, meaning that there is no radial flow of material. In a explicit form, 
\begin{align}
    q^\mu &= \big(0, q, 0, 0\big), \\
    u^\nu &= \left(\frac{1+f(t,r)}{1-f(t,r)}, 0, 0, 0\right), \\ 
    q^\mu u_\mu &=0. 
\end{align}
and where 
\begin{equation}
    f(t, r) = \frac{G\,m(t)}{2c^2r}.
\end{equation}

Replacing the expression for the imperfect fluid in the Einstein Equations yield a system of three equations for five unknowns: $a(t)$, $m(t)$, $\rho(t,r)$, $P(t, r)$ and $q$. To solve it, the functions $a(t)$ and $m(t)$ may be specified to determine the rest of the functions. The system reads
\begin{align}
    &\frac{\dot{a}(t)}{a(t)} + \frac{\dot{m}(t)}{m(t)} = -\frac{4\pi r^2a^2(t)}{m(t)}\big(1-f(t,r)\big)^2\big(1+f(t,r)\big)^4q, \\
     &\rho(t, r) = \frac{3c^2}{8\pi G}\frac{\big(1+f(t,r)\big)^2}{\big(1-f(t,r)\big)^2}\,g^2(t,r), \\
    & P(t,r) = -\frac{c^2}{8\pi G}\frac{\big(1+f(t,r)\big)^2}{\big(1-f(t,r)\big)^2}\,h(t,r)
\end{align}
where
\begin{align}
    g(t,r) &=  \frac{\dot{a}(t)}{a(t)}+\frac{G}{c^2r}\frac{\dot{m}(t)}{1+f(t,r)}, \\
    h(t,r) &=2\dot{g}(t,r)+3g^2(t,r)+\frac{2G\dot{m}(t)\,g(t,r)}{c^2r\big(1+f(t,r)\big)\big(1-f(t,r)\big)}.
\end{align}

There is a particular kind of Generalized McVittie solution which corresponds to the choice
\begin{equation}
    m(t) = m_0, 
\end{equation}
where $m_0$ is a positive constant. In isotropic coordinates, this corresponding line element is
\begin{align}
        \text{d}s^2 &= -c^2\left(\frac{1-\dfrac{G\,m_0}{2c^2r}}{1+\dfrac{G\,m_0}{2c^2r}}\right)^2 \text{d}t^2 + \\
        &+a^2(t)\left(1+\frac{G\,m_0}{2c^2r}\right)^4\big(\text{d}r^2+r^2\text{d}\theta^2+r^2\sin^2\theta\text{d}\varphi^2\big). \nonumber
\end{align} 
In the limit $m_0\to 0$ the FLRW metric is recovered, while if $a(t)=1$ it reduces to the Schwarzschild solution. 

The Misner-Sharp-Hernández (MSH) energy -- also called mass -- provides a measure of the local energy of a system in spherically symmetric spacetimes \citep{misner_sharp_1964} 
\begin{equation}\label{MSH_explicit}
    E =-\frac{1}{2}R^3K,
\end{equation}
where $R$ is the areal radius and $K$ the extrinsic curvature, both real valued functions on spacetime \citep{carrera_2010}. This is the purely geometric definition in terms of the Riemann curvature and it allows us to decompose it into the sum of two terms that come from the Ricci tensor and the Weyl curvature. It is the latter that is identified with the gravitational mass of the central object in the Newtonian limit. If $E_R$ is the part of MSH related to the Ricci tensor and $E_W$ to the Weyl curvature, then 
\begin{align}
    E_R &= \frac{4\pi}{3}R^3\rho(t,r), \\
    E_W &= a(t)\,m(t)c^2.
\end{align}
The Weyl part of the MSH energy is spatially constant and may depend on time. If that is the case then the central mass interchanges energy with the background matter. Note that the Ricci part of the MSH energy is locally related to the matter's energy momentum tensor of the components of the background. 

For the specific choice $m(t) =m_0$, the MSH energy reads 
\begin{equation}\label{msh_energy}
    E_W = c^2m_0\,a(t).
\end{equation}
From the expression above, the quantity $M(t)$ can be defined as
\begin{equation}\label{mass_change_mcvittie}
    M(t): = m_0\, a(t)
\end{equation}
and is interpreted as the effective gravitational mass of the central object. It is coupled to the dynamics of the background spacetime. 

As mentioned in the introduction, \cite{per+22} showed that a cosmological black hole, represented by the Generalized McVittie metric, survives a cosmological bounce. The gravitational mass of the black hole grows or diminishes depending on the dynamical state of the cosmological background. 

In this work, we consider that each black hole composing the population is coupled to the background dynamics, following Eq. \eqref{mass_change_mcvittie}. We also model the black hole population as a dust-like fluid. 

A cosmological dust-like fluid is a pressureless fluid, i.e., $P =0$, in which its components do not interact. In our case, we assume that during the different cosmological epochs (contraction, bounce, and expansion), the black holes do not merge. This assumption is supported by the calculations presented below.

\section{Black hole gas}\label{sec4}

\subsection{Filling factor}

The filling factor, $\mathcal{F}(\eta)$, is defined as the ratio of the area of the black hole event horizon, $A_\text{BH}$, to the area of a space-like spherical hypersurface in the FLRW geometry, $\Sigma_\text{hyp}$:
\begin{equation}
      \mathcal{F}(\eta) = \frac{A_\text{BH}(\eta)}{\Sigma_\text{hyp}(\eta)}. 
\end{equation}
We assume that there is one black hole per unit hypersurface. If both the hypersurface area and the event horizon area increase (or decrease) at the same rate, the filling factor remains constant, and therefore no black hole mergers occur. If the filling factor decreases with time, there is still one black hole per unit hypersurface, and we again infer that no mergers take place. However, if the event horizon area grows faster than the hypersurface, we assume that mergers will occur.

Although the coupling between the black holes and the cosmological background is motivated through the generalized McVittie spacetime discussed in Sec. \ref{sec3}, in the following, we approximate each individual object of the black hole population as a Schwarzschild-like black hole with an effective gravitational mass evolving in time. In other words, we adopt a quasi-local approximation in which the black hole horizon is characterized by the instantaneous Schwarzschild radius associated with the effective mass $M(\eta)$:
\begin{equation}\label{area_bh}
A_\text{BH}(\eta)  = \frac{16\pi G^2}{c^4} M^2(\eta).
\end{equation}
This mass may vary due to several processes. In this work, we consider three such processes: accretion of background radiation, evaporation via Hawking radiation, and the coupling with the background cosmological spacetime. These effects can be expressed as follows. 
\begin{equation}\label{mass_variation}
    \frac{{\rm d} M}{{\rm d}\eta} = \frac{{\rm d}M_\text{ac}}{{\rm d}\eta}+\frac{{\rm d}M_\text{RH}}{{\rm d}\eta}+\frac{{\rm d}M_\text{b}}{d\eta}.
\end{equation}

\subsubsection{Coupling with background dynamics}

As we discussed in Sec. \ref{sec3}, \cite{per+22} showed that the MSH energy corresponding to the generalized McVittie spacetime is 
\begin{equation}
    M(t) = m_0a(t). 
\end{equation}
Consequently, the mass variation is:
\begin{equation}\label{dinamica}
    \frac{{\rm d}M_\text{b}(\eta)}{{\rm d}\eta} = m_0 \frac{{\rm d}a(\eta)}{{\rm d}\eta}. 
\end{equation}

\noindent Here, $m_0$ is the mass of the black hole at $\eta = \eta_i$. Integrating Eq. (\ref{dinamica}) yields 
\begin{equation}\label{var_masa_dinamica}
     M_\text{b}(\eta) = m_0\,a_b\Bigg[1+\left(\frac{\eta}{\eta_b}\right)^2\Bigg]^{1/2}, 
\end{equation}
and the normalized mass variation is simply 
\begin{equation}
   \frac{M_\text{b}(\eta)}{m_0} = a_b\Bigg[1+\left(\frac{\eta}{\eta_b}\right)^2\Bigg]^{1/2} = a(\eta). 
\end{equation}
This function allows us to study the order of magnitude of the mass variation independently of the initial mass. 

The scale factor evolves from $a(0)=a_b = 7.41\times10^{-9}$ to $a(\pm\eta_i)=1$, implying that the black hole mass is modified by nine orders of magnitude, independently of the initial mass. 

\begin{figure}
    \centering
    \includegraphics[width=\linewidth]{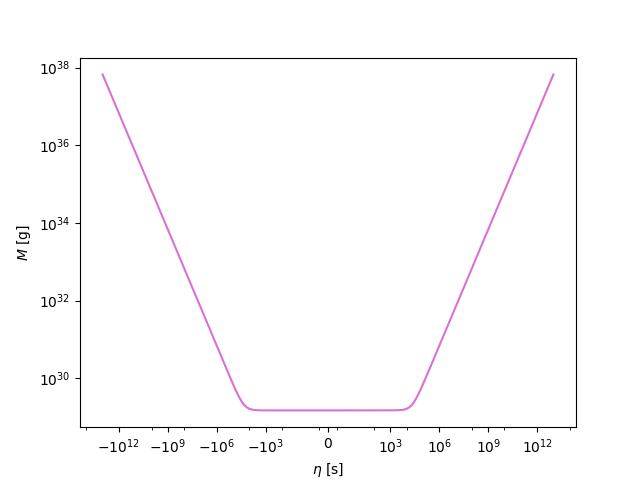}
    \caption{Mass variation for a black hole with initial mass $M=10^4 M_\odot$ due to the coupling with the background dynamics as a function of the conformal time.}
    \label{fig:var_masa_dinamica}
\end{figure}

Figure \ref{fig:var_masa_dinamica} shows the mass variation for a black hole with initial mass $M(-\eta_i) = m_0 = 10^4 M_\odot$. As expected, in the contracting phase the black hole mass decreases, reaches a minimum at the bounce and increases in the expanding phase.

\subsubsection{Hawking radiation}

In 1974, Stephen Hawking, using arguments from quantum field theory, proposed that Schwarzschild black holes emit thermal radiation in the regions near the event horizon \citep{hawking_1974}. It is therefore possible to associate a temperature with the black hole horizon:
\begin{equation}
        T = \frac{\hbar c^3}{8\pi\, G k_\text{B}\,M}.
\end{equation}
Since the temperature is inversely proportional to the mass, smaller black holes are significantly hotter and therefore lose mass more rapidly. 
The mass variation rate for Hawking radiation, expressed in terms of conformal time, is given by \citep{macgibbon_carr_1991}:
\begin{equation}\label{hawking}
    \frac{{\rm d}M_\text{RH}(\eta)}{{\rm d}\eta} = - a(\eta)\,\frac{A(M)}{M^2(\eta)},
\end{equation}
where $A(M)$ is a constant that depends on the black hole mass:
\begin{equation}
    A(M) = \left\{ \begin{array}{rcl}
    & 5.3\times10^{25}~\mathrm{g}^3\,\mathrm{s}^{-1}~ ~\text{if}~~M> 10^{17}~\mathrm{g}, \\  
    & 7.8\times10^{27}~\mathrm{g}^3\,\mathrm{s}^{-1}~~\text{if}~~M\leq 10^{15}~\mathrm{g}. 
\end{array}\right.        
\end{equation}

\begin{figure}
    \centering
    \includegraphics[width=\linewidth]{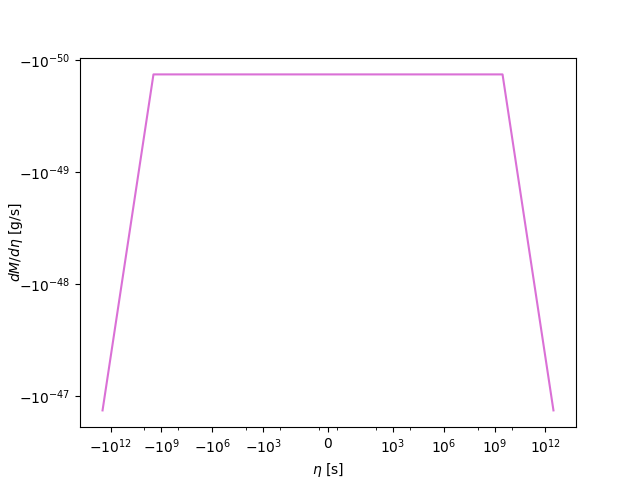}
    \caption{Rate of mass variation  d$M_\text{RH}/$d$\eta$ for a black hole with initial mass $M = 10^4M_\odot$ due to Hawking radiation as a function of the conformal time.}
    \label{fig:var_masa_hawking}
\end{figure}

As the rate of mass variation (\ref{hawking}) is proportional to $M^{-2}$, this effect can be neglected for massive black holes. In Fig. \ref{fig:var_masa_hawking} we show the  rate of mass variation for a black hole with mass $M_\text{inic} = 10^4M_\odot$ at time $\eta=-\eta_i$. On the other hand, less massive black holes ($M<10^{15}$ g) evaporate before the bounce occurs. 

In this work, we focus on black holes that persist throughout the three cosmological epochs (contraction, bounce, and expansion). Accordingly, we restrict our analysis to black holes with masses $M > 10^{16}$ g, for which the mass loss rate due to Hawking radiation is negligible relative to that associated with the background dynamics.

\subsubsection{Accretion of the cosmological fluid}\label{rad_acc}

The black hole mass may increase through accretion of the background radiation. The mass variation associated with this effect is given by \citep{zeldovich1967}:
\begin{equation}\label{var_masa}
    \frac{{\rm d}M}{{\rm d}\eta} = \frac{27\pi G^2}{c^3}\, a(\eta)\, \rho_\text{rad}\, M^2,
\end{equation}
where $\rho_\text{rad}$ is the radiation density. From the Friedmann equation in terms of the conformal time
\begin{equation}
    \left(\frac{a'(\eta)}{a^2(\eta)}\right)^2 = \frac{8\pi G}{3}\,\rho_\text{rad}(\eta),
\end{equation}
and replacing the scale factor given by Eq. \eqref{factor_de_escala}, we obtain:
\begin{equation}\label{densidad_radiacion}
    \rho_\text{rad}(\eta) = \frac{3}{8\pi G}\frac{1}{\eta_b^4a_b^2}~\eta^2\left[1+\left(\frac{\eta}{\eta_b}\right)^2\right]^{-3}.
\end{equation}
For $\eta\gg 1$: 
\begin{equation}
     \rho_\text{rad}(\eta) \sim \left(\frac{\eta}{\eta_b}\right)^{-4} \sim a^{-4}(\eta), 
\end{equation}
as expected far away from the bounce. 

Equation \eqref{var_masa} takes the following form:
\begin{equation}
    \frac{{\rm d}M}{{\rm d}\eta} = \frac{81}{8}\frac{G}{c^3} \frac{1}{\eta_b^2\,a_b}\frac{\eta^2}{\eta_b^2}\left[1+\left(\frac{\eta}{\eta_b}\right)^2\right]^{-5/2}\,M^2(\eta).
\end{equation}
Introducing the dimensionless variable $x = \eta/\eta_b$, and a constant $\sigma$ defined by:
\begin{equation}
    \sigma = \frac{81}{8}\frac{G}{c^3} \frac{1}{\eta_b^2\,a_b},
\end{equation}
we find:
\begin{equation}
    \frac{{\rm d}M}{{\rm d}x} = \sigma\,x^2\left[1+x^2\right]^{-5/2}\,M^2(\eta).
\end{equation}

We solve this differential equation using the boundary condition $M_\text{bc}=M_\text{bc}(x_\text{bc})$, where $x_\text{bc} = \eta_\text{i}/\eta_\text{b}$. The integration leads to 
\begin{equation}\label{mass_acre}
    M(x) = \left(-\frac{\sigma\, x^3}{3\big(x^2+1\big)^{3/2}}+g_\text{bc}\right)^{-1},
\end{equation}
where we have defined
\begin{equation}
    g_\text{bc} = \frac{\sigma\,x^3_\text{bc}}{3\big(x_\text{bc}^2+1\big)^{3/2}}+\frac{1}{M_\text{bc}}.
\end{equation}
We rewrite \eqref{mass_acre} in terms of the normalized mass function $M(x)/M_\text{bc}$:
\begin{equation}\label{masa_norm_acrecion}
    \frac{M(x)}{M_\text{bc}} = \left(-\frac{\sigma\, M_\text{bc}\, x^3}{3\big(x^2+1\big)^{3/2}}+g'_\text{bc}\right)^{-1},
\end{equation}
with 
\begin{equation}
    g'_\text{bc} =  g_\text{bc}\,M_\text{bc}.
\end{equation}

\begin{figure}
    \centering
    \includegraphics[width=\linewidth]{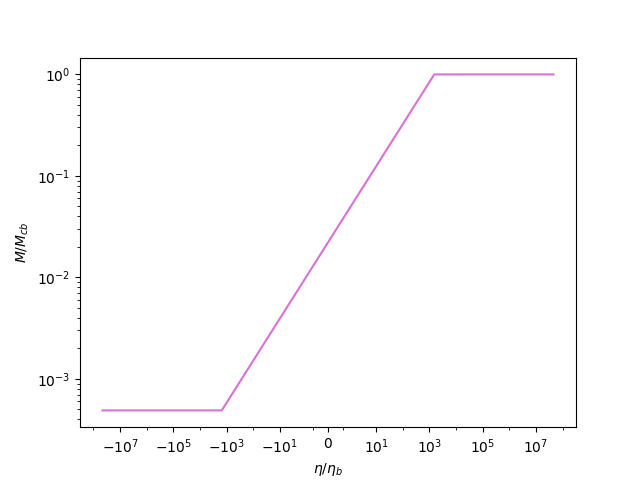}
    \caption{Mass variation for a black hole with final mass $M_\text{bc}(\eta_i) = 10^4M_\odot$ due to the accretion of a radiation as a function of the normalized conformal time $\eta/\eta_b$.}
    \label{fig:var_masa_acrecion}
\end{figure}

Figure \ref{fig:var_masa_acrecion} shows the normalized mass function considering that the black hole reaches a final mass of $M_\text{bc}(\eta_i) = 10^4M_\odot$. The mass starts to increase considerably at $\eta/\eta_b = -10^3$, which corresponds to $\eta=-2.9\times10^7$ s. 

Depending on the value of the mass at the boundary condition, that is $M(x_\text{bc})$, the value of the mass at $M(-x_\text{bc})$ changes. We present three different cases: 
\begin{itemize}
    \item If $M_\text{bc} = 10^4M_\odot$, $M(-x_\text{bc})/M_\text{bc} \sim 10^{-4}$. 
    \item If $M_\text{bc} = 10^6M_\odot$, $M(-x_\text{bc})/M_\text{bc} \sim 10^{-6}$.
    \item If $M_\text{bc} = 10^8M_\odot$, $M(-x_\text{bc})/M_\text{bc} \sim 10^{-8}$.
\end{itemize}
The more massive the black hole at $x_\text{bc}$, the greater the mass variation within the interval $(-x_\text{bc}, x_\text{bc})$. For black holes with masses below $10^{8}\,M_{\odot}$, the variation due to accretion is at least an order of magnitude smaller than that produced by coupling to the background dynamics. Recall that the latter effect is independent of mass and consistently amounts to nine orders of magnitude. In this work, we therefore restrict our analysis to black holes in the mass range $10^{16}$ g $< M < 10^6 M_\odot$, for which coupling with the cosmological dynamics is the only mechanism that significantly alters the mass.

\subsubsection{Filling factor: results}

Since in the mass range $10^{16}$ g $<M<10^6M_\odot$, the mass variation is essentially dominated by the coupling with the background cosmological dynamics, the area of the black hole event horizon yields
\begin{equation}\label{area_bh_fescala}
    A_\text{BH}(\eta) = \frac{16\pi\,G^2\,m_0{^2}}{c^4}   ~a^2(\eta).
\end{equation}

On the other hand, the area of a space-like hypersurface from the FLRW line element ,
\begin{equation}
ds^{2} = a^{2}(\eta) \left(-c^2 d\eta^2 + dr^2 + r^2  \; d\theta^2 + r^2  \;\sin^{2}{\theta} \; d\phi^2 \right),
\end{equation}
takes the form
\begin{equation}\label{area_hipsup}
 \Sigma_\text{hyp}(\eta) =4\pi r^2\,a^2(\eta).
\end{equation}

Replacing  Eqs. \eqref{area_bh_fescala} and \eqref{area_hipsup}, the filling factor results in: 
\begin{equation}
    \mathcal{F}(\eta) = \frac{A_\text{BH}(\eta)}{\Sigma_\text{hyp}(\eta)} = \frac{16\pi\,m_0{^2}\,G^2~a^2(\eta)}{4\pi c^4\,r^2~a^2(\eta)} = \frac{4m_0{^2}\,G^2}{c^4\,r^2}.
\end{equation}
The filling factor does not depend on the conformal time. We therefore deduce that the area of the event horizon evolves in the same way as that of the hypersurface. Since we assume there is only one black hole per unit hypersurface, and both areas increase or decrease at the same rate, no two black holes will ever occupy the same hypersurface. Under these assumptions, we conclude black holes do not merge.

\subsection{Gravitational radiation}

Gravitational radiation is another mechanism that can lead to black hole mergers. During the contraction phase of the universe, two nearby black holes may form a binary system and begin orbiting their center of mass. As they lose energy through gravitational radiation, their separation decreases until a merger becomes possible. If the merger timescale is shorter than the time to the bounce, we infer that the black holes will indeed merge.

We consider two point-like masses $M_1$ and $M_2$ in a circular orbit of radius $\zeta$. The fusion time is given by  \citep{Shapiro_Teukolsky_book}:
\begin{equation}\label{eq:tiempo_fusion}
    t_\text{fus} = \frac{5}{256}\frac{c^5}{G^3\mu M^2}~\zeta^4, 
\end{equation}
where $\mu$ is the reduced mass 
\begin{equation}
    \mu = \frac{M_1M_2}{M_1+M_2},
\end{equation}
and $M = M_1+M_2$. Note that the physical distance between the black holes $d =2\zeta$ depends on the scale factor. If $d_0$ is the distance at the conformal time $\eta_i$, the expression for the physical distance as a function of the conformal time is
\begin{equation}
    d(\eta) = d_0\,a(\eta). 
\end{equation}

We now compute $t_\text{fus}$ for a binary system of supermassive black holes of initial mass $M_1 = M_2 = 10^6 M_\odot$; we take into account that their masses change due to the coupling with the background dynamics as given by Eq. \eqref{var_masa_dinamica}. The initial distance between them is $d_0 = 2.36\times10^{24}$ cm, which is equivalent to the distance between the Milky Way and the Andromeda galaxy. 

\begin{figure}
    \centering
    \includegraphics[width=\linewidth]{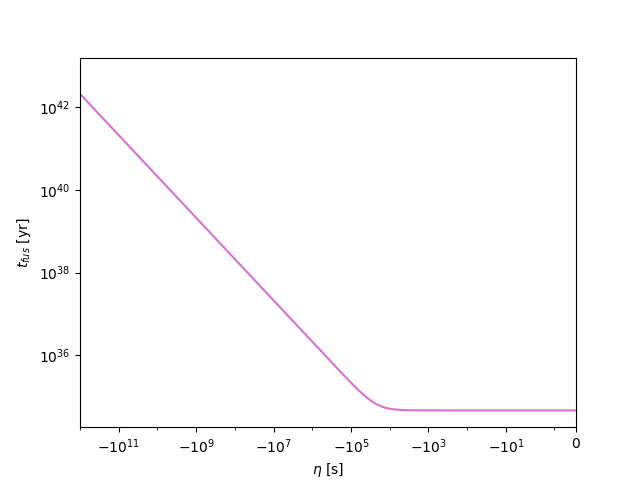}
    \caption{Merging time, in years, in the contraction phase for two black holes with initial mass $M_1=M_2=10^6 M_\odot$ and an initial distance $d_0 = 2.36\times10^{24}$ cm at $\eta_i = -2.96\times10^{12}$ s as a function of the conformal time.}
    \label{fig:tiempo_fusion}
\end{figure}

Figure \ref{fig:tiempo_fusion} shows the merger time, in years, during the contraction phase. It is always longer than the time to the bounce (recall that the radiation epoch lasts approximately $47{,}000$ years \citep{Ryden_2016}). Had we chosen smaller binary masses, the merger time would be even longer, since $t_\text{fus}$ is inversely proportional to the cube of the mass. We therefore conclude that, under the assumptions adopted, no mergers between black holes occur.

In summary, we have shown that, under certain assumptions, black holes do not undergo mergers. Consequently, we model the population as a dust-like fluid, setting $P_\text{BH}=0$.

\section{Interacting fluids}\label{sec5}

We consider a perfect fluid immersed in a FLRW background. The corresponding energy-momentum tensor is: 
\begin{equation}\label{tmunu}
    T_{\mu\nu} = (\rho+P)u_\mu u_\nu +Pg_{\mu\nu},
\end{equation}
where $P$ is the pressure, $\rho$ is the energy density and $u^\mu$ the four-velocity of the fluid. 

The fluid has two components: a) $T^{\mu\nu}_\text{BH}$, related to the population of black holes and, b) $T^{\mu\nu}_\text{CF}$, associated with the background cosmological fluid. These components interchange energy and momentum, implying that only the total energy-momentum tensor is conserved:
\begin{equation}\label{cons_parical}
    \nabla_\mu T^{\mu\nu}_\text{total} =  \nabla_\mu(T^{\mu\nu}_\text{BH} + T^{\mu\nu}_\text{CF}) =0. 
\end{equation}

This leads to:

\begin{equation}
\nabla_\mu T^{\mu\nu}_\text{BH}=-\nabla_\mu T^{\mu\nu}_\text{CF}.
\end{equation}

 From Eq. \ref{cons_parical}, we find:
\begin{eqnarray}\label{casi_sistema_a_resolver}
    \rho_\text{BH}'(\eta) + 3\frac{a'(\eta)}{a(\eta)}\big(\rho_\text{BH}(\eta)+P_\text{BH}(\eta)\big) & = & ~ Q(\eta), \label{casi_sistema_a_resolver1}\\
    \rho_\text{CF}'(\eta) + 3\frac{a'(\eta)}{a(\eta)}\big(\rho_\text{CF}(\eta)+P_\text{CF}(\eta)\big)& = &- Q(\eta),\label{casi_sistema_a_resolver2}
\end{eqnarray}
where $Q$ is the interaction term. 

The pressure depends on the properties of the fluid. Since one of the goals of this work is to characterize the evolution of the background cosmological fluid in the presence of black holes, $P_\text{CF}(\eta)$ is one of the unknown functions to determine. As we found in Sect. \ref{sec4}, $P_\text{BH}=0$ if the black hole population is in the mass range $10^{16}$ g $< M < 10^6 M_\odot$. 

The system of Eqs. \eqref{casi_sistema_a_resolver1}  and \eqref{casi_sistema_a_resolver2} has three unknowns: $\rho_\text{BH}$, $\rho_\text{CF}$ and $P_\text{CF}$. The system is completed by adding one of the Friedmann equations expressed in terms of the conformal time ($k=\Lambda=0$): 
\begin{equation}\label{friedmann}
\left(\frac{a'(\eta)}{a^2(\eta)}\right)^2 =  \frac{8\pi G}{3c^2} \rho(\eta),
\end{equation}
where $\rho(\eta) = \rho_\text{BH}(\eta)+\rho_\text{CF}(\eta)$. 

We characterize the black holes that constitute the dust-like fluid by their initial mass function (IMF) $N_0(m)$, since they may span a range of masses. This function evolves in time due to its coupling with the background dynamics and the energy exchange between each black hole and the cosmological fluid. If $N(\eta; m) $ is the number density of the black holes (i.e., the mass function at conformal time $\eta$), the energy density of the black hole gas can be written as \citep{Borunda_2010, barrow1991}:
\begin{equation}\label{densida_bh_integral}
    \rho_{\rm BH}(\eta) = \int^{M_\text{max}}_{M_\text{min}} N(\eta; m) E(m)~ {\rm d}m.
\end{equation}
Here, $E(m) = mc^2$ denotes the energy of a Schwarzschild black hole, while $M_\text{min}$ and $M_\text{max}$ represent the minimum and maximum black hole masses, respectively. The integration is performed at constant conformal time.

The number density of black holes changes due to the expansion of the volume and the evolution of the mass. The continuity equation is a Boltzmann equation and is of the form 
\begin{equation}\label{eq:continuity1}
\frac{\partial N}{\partial \eta} + 3\mathcal{H} N + \frac{\partial}{\partial m} \left(N \frac{\partial m}{\partial \eta}\right) = S(m,\eta)
\end{equation}
where $\mathcal{H}(\eta) = a'/a$ is the Hubble parameter in terms of the conformal time, $3\mathcal{H}(t)N$ accounts for the volumetric dilution, $\partial (Nm')/\partial m$ is the "flow" in mass space (advection) and $S(m,\eta)$ represents any source/sink terms (mergers or evaporation), which we consider $S(m,\eta)=0$. 

As we only consider that the mass change is due to the coupling to the dynamics of spacetime, 
\begin{equation}
    m'= m_0 a' = m_0a\frac{a'}{a} = m(\eta)\mathcal{H}(\eta). 
\end{equation}
This allows us to write the continuity equation for $N(m, \eta)$ as 
\begin{equation}\label{eq:boltzmann}
    \frac{\partial N}{\partial \eta} + 3\mathcal{H} N + \mathcal{H}\frac{\partial \left(N m\right)}{\partial m}  = 0.
\end{equation}

To find the evolution of $\rho_{\rm BH}(\eta)$ we multiply Eq. \eqref{eq:boltzmann} by $mc^2$ and integrate over the mass spectrum. 


We integrate by parts the right hand side
\begin{equation}
     \int \frac{\partial \left(N m\right)}{\partial m} \,mc^2~{\rm d}m = c^2m^2N\Big|^{M_\text{max}}_{M_\text{min}}\infty_0 
     - \int N\, mc^2~ {\rm d}m,
\end{equation}
and we assume that the distribution $N(m,\eta)$ decays sufficiently fast at large masses and does not diverge strongly at small masses, such that $m^2 N(m,\eta) \rightarrow 0$ at the integration boundaries. Under this assumption, the boundary term vanishes. This is equivalent to assuming no flux of objects across the mass boundaries. For instance, this condition is automatically satisfied for a monochromatic distribution $N(m,\eta) \propto \delta(m-m_0)$, provided that $m_0$ lies within the integration range, and also for power-law distributions $N(m,\eta) \propto m^{-\alpha}$ with suitable cutoffs and indices such that $m^{2-\alpha} \rightarrow 0$ at the boundaries.

 Therefore, 
  \begin{equation}
      \frac{\partial \rho_{\rm BH}}{\partial \eta} +3\mathcal{H}\rho_{\rm BH} =\mathcal{H} \int N\, mc^2~ {\rm d}m=\mathcal{H}\rho_{\rm BH}.
 \end{equation}

If we compare the later equation with Eq. \eqref{casi_sistema_a_resolver1}, we identify the exchange of energy as 
\begin{equation}
    Q(\eta) = \mathcal{H}\rho_{\rm BH}.
\end{equation}
The solution of the energy density for the black hole population is  
\begin{equation}\label{evol_dens_bh}
     \frac{\partial \rho_{\rm BH}}{\partial \eta} +2\mathcal{H}\rho_{\rm BH} = 0 \implies \rho_{\rm BH}=  \rho_{\rm BH}(\eta_i)\left(\frac{a(\eta_i)}{a(\eta)}\right)^2,
\end{equation}
where $a(\eta_i)=1$ and $\rho_\text{BH}(\eta_i)$ is the energy density of the black holes in the transition time.

It is seen that, macroscopically, black holes coupled to the background dynamics (whose mass is proportional to the scale factor) dilutes slower than non-relativistic matter. This scaling mimics the behavior of spatial curvature. However, near the bounce, it will not dominate over radiation, which scales as $\rho\propto a^{-4}$. 

Another important result is that the evolution of the energy density of the black holes do not depend on the number density. Any kind of population will evolve in the same way, regardless of its distribution. 

Once we found the evolution of the black hole gas, we use the Friedmann equation \eqref{friedmann} to solve for the cosmological fluid 
\begin{equation}\label{evol_dens_cf}
    \rho_\text{CF}(\eta) =  \frac{3c^2}{8\pi\, G}\left(\frac{a'(\eta)}{a^2(\eta)}\right)^2 - \rho_\text{BH}(\eta). 
\end{equation}

One can solve Eq. \eqref{casi_sistema_a_resolver2} for the pressure of the cosmological fluid. Another approach is using the Raychaudhuri equation
\begin{equation}
         \frac{a''(\eta)}{a^3(\eta)} =\frac{4\pi G}{3c^2}  \big(\rho_\text{BH} + \rho_{CF} - 3P_\text{CF}\big).
\end{equation}
Replacing the explicit expressions we find 
\begin{equation}\label{pressure}
   P_\text{CF} =  \frac{c^2}{8\pi \, G}\left(\frac{a'(\eta)}{a^2(\eta)}\right)^2 -\frac{c^2}{4\pi \, G}  \frac{a''(\eta)}{a^3(\eta)}. 
\end{equation}

Once specified the scale factor $a(\eta)$ we find the explicit results for the evolution of the black holes, the cosmological fluid and pressure of the cosmological fluid. It is worth mentioning that as we fix the geometry of the spacetime we are interested in investigating how the properties of the cosmological fluid change in the phases of contraction, bounce and expansion considering the presence of the black holes. The analysis of the evolution of the cosmological fluid in absence of the black hole population is found in the \ref{app2}.

\section{Results}\label{sec6}

The spacetime geometry is fixed for the scale factor given in Eq. \eqref{factor_de_escala}. If we replace it in the solutions \eqref{evol_dens_bh}, \eqref{evol_dens_cf} and \eqref{pressure}, we find 
\begin{eqnarray}
    \rho_\text{BH}(\eta) &= &\frac{a^2(\eta_i)\rho_\text{BH}(\eta_i)}{a_b(\eta_b^2+\eta^2)}, \\
     \rho_\text{CF}(\eta) &= &\frac{3c^2\eta_b^2\,\eta^2}{8\pi Ga_b^2(\eta_b^2+\eta^2)^3} - \frac{\eta_b^2a^2(\eta_i) \rho_\text{BH}(\eta_i)}{a_b^2(\eta_b^2+\eta^2)}, \\
     P_\text{CF} (\eta) & =& \frac{c^2\eta_b^2\,\eta^2}{8\pi G\,a_b^2(\eta_b^2+\eta^2)^3} - \frac{2c^2\eta_b^4}{8\pi G\,a_b^2(\eta_b^2+\eta^2)^3}.
\end{eqnarray}
We impose the boundary conditions at the conformal time $\eta_\text{i} = 2.96\times10^{12}$ s, which corresponds to the transition time from the radiation-dominated epoch to the matter-dominated epoch. 

We assume that, at this conformal time, the energy density of the black hole gas is a fraction of the radiation energy density
\begin{equation}
    \rho_\text{BH}(\eta_\text{i}) = \beta\,\rho_\text{rad}(\eta_\text{i}),
\end{equation}
where $\beta$ is a free parameter of the model, which restricts the initial abundances of primordial black holes. Although the black hole population considered here is not necessarily of primordial origin, we adopt the same bounds on $\beta$ to ensure consistency with the sensitivity of current observational constraints.


Primordial black holes with masses $M_\text{PBH} >10^{15}$ g could survive until the present epoch and be detected through their gravitational effects, in particular via their contribution to a stochastic gravitational wave background. In the case of massive primordial black hole binaries that may be merging today and producing gravitational waves, current LIGO observations place only weak restrictions. Following \citep{Carr2010}, the abundance parameter is constrained to $\beta<10^{-7}$ for stellar black holes. Throughout this work we adopt a value $\beta=10^{-10}$. 

From Eq. \eqref{densidad_radiacion} we calculate the energy density of radiation at the boundary condition, which yields $\rho_\text{rad}(\eta_\text{i}) = \rho_\text{CF}(\eta_\text{i})=184.35$ erg\,cm$^{-3}$. Then, the energy density for the black hole gas at $\eta_\text{i}$ is
\begin{equation}
    \rho_\text{BH}(\eta_\text{i}) =  1.84\times10^{-8}~\text{erg cm}^{-3}.
\end{equation}

\begin{figure}
    \centering
    \includegraphics[width=\linewidth]{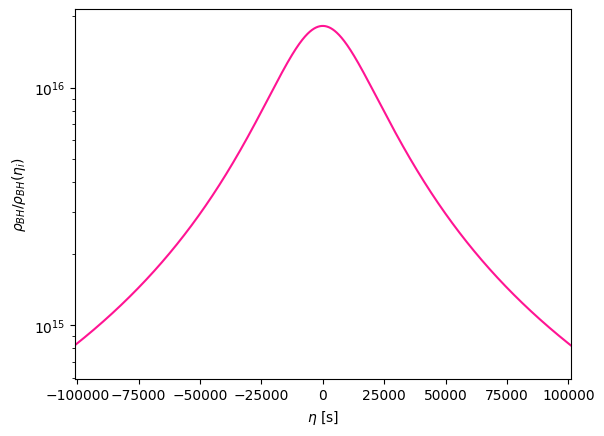}
    \caption{Evolution of the energy density for the black hole population as a function of the conformal time in the interval $-1.10\times10^5$ s $<\eta<1.10\times10^5$ s.}
    \label{fig:rho_bh_mono}
\end{figure}

Figure \ref{fig:rho_bh_mono} shows the evolution of the normalized black hole energy density $\rho_\text{BH}(\eta)/\rho_\text{BH}(\eta_\text{i})$, for the time interval $-1.10\times10^{5}$ s $<\eta<1.10\times10^{5}$ s.  The energy density increases as the bounce is approached during the contracting phase, reaching a maximum at the bounce, with a value of $3.35\times10^8$ erg\,cm$^{-3}$,  and then decreases as the universe expands.

\begin{figure}
    \centering
    \includegraphics[width=\linewidth]{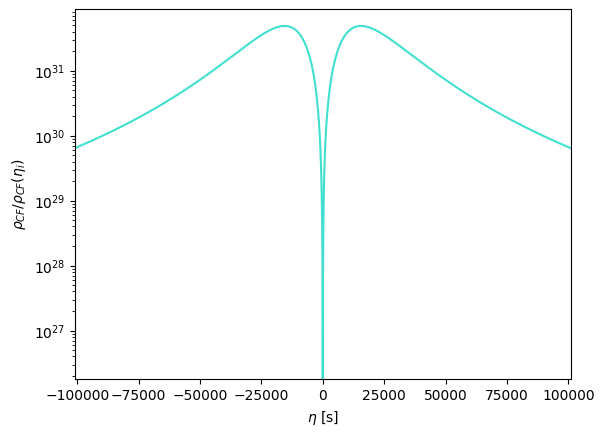}
    \caption{Evolution of the energy density for the cosmological fluid as a function of conformal time in the interval $-1.10\times10^5$ s $<\eta<1.10\times10^5$ s.}
    \label{fig:rho_cf_mono}
\end{figure}

\begin{figure}
    \centering
    \includegraphics[width=\linewidth]{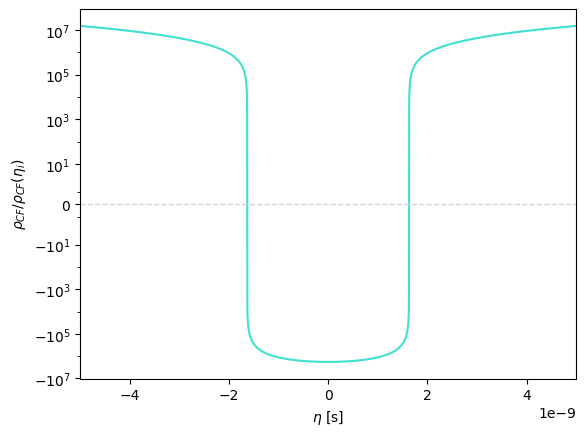}
    \caption{Evolution of the energy density for the cosmological fluid as a function of conformal time in the interval $-5\times10^{-9}$ s $<\eta<-5\times10^{-9}$ s. The dotted line corresponds to $\rho_\text{CF} = 0$.}
    \label{fig:rho_cf_mono_cerca}
\end{figure}

In Fig. \ref{fig:rho_cf_mono}, we depict the normalized energy density of the  cosmological fluid in the range $-1.10\times10^5$ s $<\eta<1.10\times10^5$s, while in Fig. \ref{fig:rho_cf_mono_cerca} we look closer to the bounce, restricting to the range $-1.10\times10^{-4}$ s $<\eta<1.10\times10^{-4}$ s. We see that the behavior of the cosmological fluid is completely symmetric with respect to the bounce. The energy density $\rho_\text{CF}(\eta)/\rho_\text{CF}(\eta_\text{i})$ has two maxima at $\eta=\pm1.55\times10^4$ s and two zeros at $\eta=\pm1.63\times10^{-9}$ s. In the interval $-1.63\times10^{-9}$ s $<\eta<1.63\times10^{-9}$ s, the energy density becomes negative. This implies that the weak energy condition (WEC) is violated. At $\eta=0$ the energy density is at its minimum and its value is  $\rho_{\rm CF}(0) = -3.34\times10^{8}$ erg\,cm$^{-3}$ $= -\rho_\text{BH}(0)$. For a bounce to happen, it must be that $a'(0) = 0$ and from Friedmann equation \eqref{friedmann} we derive the result aforementioned. The energy density of the cosmological fluid must become negative in a certain range of time to accommodate the geometry of spacetime that we have fixed. 

\begin{figure}
    \centering
    \includegraphics[width=\linewidth]{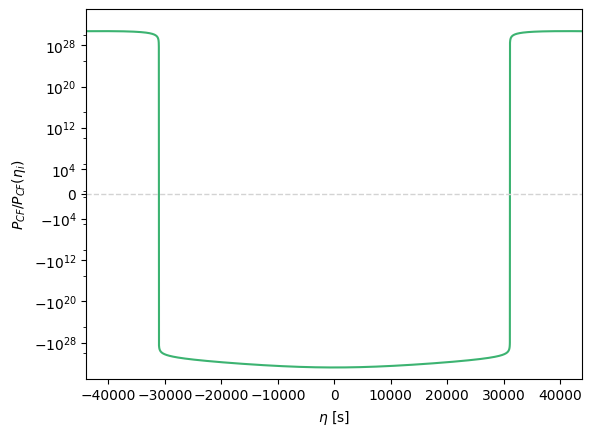}
    \caption{Evolution of the pressure of the cosmological fluid as a function of the conformal time in the interval $-4.39\times10^4$ s $<\eta<4.39\times10^4$ s. The dotted line corresponds to $P_\text{CF} = 0$.}
    \label{fig:presion}
\end{figure}

Figure \ref{fig:presion} depicts the evolution of the normalized pressure $P_\text{CF}(\eta)/P_\text{CF}(\eta_\text{i})$. In this case, the normalization is $P_\text{CF}(\eta_\text{i}) = 61.45$ erg\,cm$^{-3}$. We see that it also has two maxima at $\eta= \pm 4.11\times10^4$ s and a minimum at $\eta=0$, where the pressure has a value of $P_\text{CF}(0) = -4.05\times10^{32}$ erg\,cm$^{-3}$. Towards the bounce, the pressure becomes negative at $\eta= \pm 3.10\times10^4$ s. This implies that the cosmological fluid behaves as a repulsive fluid. 

We also show the energy conditions. As we clarified in Sec. \ref{sec2}, the NEC and the SEC do not change under the presence of the black holes since they only depend on the scale factor, which we have fixed. In explicit form, and in terms of the conformal time, the NEC is
\begin{align}
    \rho+P &= \rho_\text{BH}+\rho_\text{CF} + P_\text{CF}, \\
    &=  \frac{c^2}{2\pi\, G}\left(\frac{a'(\eta)}{a^2(\eta)}\right)^2  -\frac{c^2}{4\pi \, G}  \frac{a''(\eta)}{a^3(\eta)},\\
    &= \frac{c^2\eta^2_b\big(2\eta^2-\eta_b^2\big)}{4\pi\,G\,a_b^2\big(\eta^2+\eta_b^2\big)^3},
\end{align}
and the SEC has the form 
\begin{align}
    \rho+3P &= \rho_\text{BH}+\rho_\text{CF} + 3P_\text{CF}, \\
    &=  \frac{3c^2}{4\pi\, G}\left(\frac{a'(\eta)}{a^2(\eta)}\right)^2  -\frac{3c^2}{4\pi \, G}  \frac{a''(\eta)}{a^3(\eta)},\\
    &= \frac{3c^2\eta^2_b\big(\eta^2-\eta_b^2\big)}{4\pi\,G\,a_b^2\big(\eta^2+\eta_b^2\big)^3}.
\end{align}

\begin{figure}
    \centering
    \includegraphics[width=\linewidth]{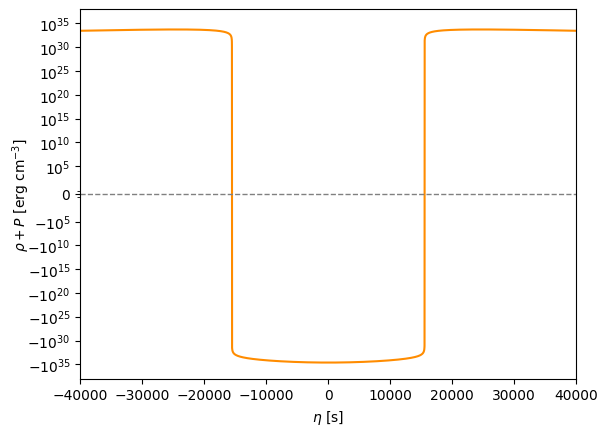}
    \caption{Plot of the NEC as a function of the normalized conformal time in the interval $-4\times10^{4}$ s $<\eta<4\times10^{4}$ s for a narrow mass function. The dotted line corresponds to $\rho + P=0$.}
    \label{fig:nec-mono}
\end{figure}

\begin{figure}
    \centering
    \includegraphics[width=\linewidth]{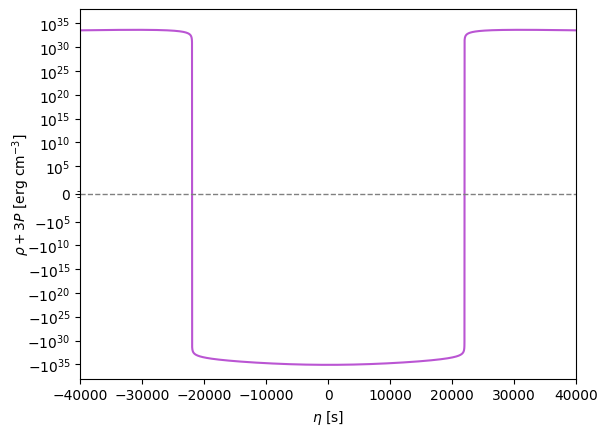}
    \caption{Plot of the SEC as a function of the normalized conformal time in the interval $-4\times10^{4}$ s $<\eta<4\times10^{4}$ s for a narrow mass function.  The dotted line corresponds to $\rho + 3P=0$.}
    \label{fig:sec-mono}
\end{figure}

As a function of the conformal time, the NEC and the SEC are depicted in Fig. \ref{fig:nec-mono} and Fig. \ref{fig:sec-mono}, respectively. The NEC is violated for $\eta \in (-1.55\times10^{4} \;\mathrm{s} , ~1.55\times10^{4} \;  \mathrm{s})$, which corresponds to the time interval between the two maxima of the energy density of the cosmological fluid. On the other hand, the SEC is violated for $\eta \in (-2.19\times10^{4}\;  \mathrm{s}, ~2.19\times10^{4} \; \mathrm{s})$ which is within the scale of the bounce.

\section{Discussion and conclusions}\label{sec7}

In this work, we have investigated the evolution of a population of black holes in a bouncing universe, taking into account their interaction with the background cosmological fluid. We have also analyzed how the cosmological fluid behaves in a spacetime with a fixed geometry. 

First, we explored whether black holes could merge through two different channels. On the one hand, we assumed that at the beginning of the contraction phase there is one black hole per unit spatial hypersurface embedded in the FLRW background. We then calculated the filling factor, defined as the ratio between the area of the event horizon and the area of a radial spacelike hypersurface in the FLRW geometry.

The area of the event horizon can change due to three possible contributions: accretion of radiation, Hawking radiation, and coupling between the event horizon and the background cosmological dynamics. Within the mass range \( 10^{16} \; \mathrm{g} < M < 10^{6} \; M_{\odot} \), the only relevant contribution is the latter. Under this condition, the filling factor remains constant in time. Therefore, we conclude that black holes do not merge.

On the other hand, if black holes come sufficiently close, particularly during the contraction phase, they may begin to spiral toward each other and eventually merge. To explore this scenario, we considered a pair of supermassive black holes (modeled as point-like masses) initially separated by a distance of 764.6 kpc, which corresponds to the average distance between the Milky Way and the Andromeda galaxy. We found that the merger timescale is significantly longer than the duration of the contraction phase. Therefore, we also neglect black hole mergers via this channel.

Assuming that black holes do not interact, we characterized the black hole population as a dust-like fluid. By imposing the conservation of the total energy-momentum tensor of the system and employing the cosmological field equations of General Relativity, we derived a system of differential equations governing the energy densities of both the black holes and the cosmological fluid, as well as the pressure of the cosmological fluid. We found that the energy density of the black holes scales as $\rho_\text{BH}\propto a^{-2}$ and does not depend on the type of distribution. Macroscopically, black holes coupled to the background dynamics dilutes slower than non-relativistic matter. 

The energy density of black holes increases significantly as the universe approaches the bounce, reaching a maximum at that point; the energy density of the cosmological fluid becomes negative around the bounce and equals the energy density of the black holes at the bounce. The cosmological fluid must alter its behavior to allow the bounce to happen, since the background geometry is fixed. 

The pressure of the cosmological fluid only depends on the form of the scale factor, due to the fact that we modeled the population of the black holes as dust. Around the bounce, the pressure becomes negative, therefore the cosmological fluid behaves as a repulsive fluid. 

The strong and null energy conditions only depend on the geometry of spacetime, which is fixed, and therefore are still violated independent of the presence of the black holes. 

There are several issues that require further investigation. First, it is important to incorporate radiation accretion for the range of black hole masses where this process is relevant. In this extended scenario, black hole mergers could occur, and the black hole population can no longer be treated as a dust-like fluid. Additionally, interactions between black holes during the contraction phase could generate a stochastic gravitational wave background, potentially leaving observable imprints on the cosmic microwave background. This remains an open area of research that has not yet been explored.

We close this article by pointing to a direction for future research: analyzing the conditions under which a population of black holes could prevent the occurrence of a cosmological bounce. In the present investigation, the bounce is unavoidable due to the specific form of the scale factor assumed. We plan to address this question in greater detail in future work.


\section*{Acknowledgments}

The authors are grateful to Gustavo E. Romero and Santiago E. Pérez Bergliaffa for many insightful discussions on this work.

\subsection*{Financial disclosure}

None reported.

\subsection*{Conflict of interest}

The authors declare no potential conflict of interests.

\appendix

\section{Model parameters}\label{app1}

The model parameters to be determined are: 
\begin{itemize}
    \item $a_b$: the normalization parameter of the scale factor.
    \item $\eta_b$: the conformal time at the scale of the bounce.
    \item $\eta_i$: the conformal time corresponding to the transition between the radiation to matter dominated epoch. 
\end{itemize}

We normalize the scale factor so that its value at the radiation-matter transition point is one:
\begin{equation}\label{eq1}
    a(\eta_i) = a_b\left[1+\left(\frac{\eta_i}{\eta_b}\right)^2\right]^{1/2} = 1.
\end{equation}

The maxima of the radiation energy density (Eq. \ref{densidad_radiacion}) are located at $\eta = \pm \eta_b/\sqrt{2}$. Evaluating Eq. \eqref{densidad_radiacion} at these points allows us to solve for $\eta_b$ at the maxima, yielding:
\begin{equation}\label{eq2}
    \eta_b = \sqrt{\frac{1}{18\pi\, G\,a_b^2\,\rho_\text{max}}},
\end{equation}
where $\rho_\text{max}=10^{13}$  g\,cm$^{-3}$. 

From the relationship between the cosmic time and conformal time (Eq. \ref{tiempo_cosmico}) evaluated at the cosmic time of transition $t_i = 47\,000$ yr, we obtain 
\begin{equation}\label{eq3}
   t_i = \frac{a_b}{2}\left[\eta_b~ \text{arcsenh}\left(\frac{\eta_i}{\eta_b}\right)+\eta_i\sqrt{\frac{\eta_i^2}{\eta^2_b}+1}\right]. 
\end{equation}

Eqs. \eqref{eq1}, \eqref{eq2} and \eqref{eq3} form a system of three equations with three unknowns. To determine the solution, we define $x =\eta_i/\eta_b$ so that the Eqs. \eqref{eq1} and \eqref{eq3} reduce to 
\begin{eqnarray}
    a_b\left[1+x^2\right]^{1/2} =&~ 1 \label{reemplazo1}, \\
    \frac{a_b\,\eta_b}{2}\left[ \text{arcsenh}\big(x\big)+x~\sqrt{x^2+1}\right] =&~t_i. \label{reemplazo2}
\end{eqnarray}
Then, we substitute the values found for $a_b$ and $\eta_b$ in Eq. \eqref{reemplazo2}. Defining a constant $\alpha^2 = 72\pi\,G\,\rho_\text{max}$, we find that Eq. \eqref{reemplazo2} takes the form
\begin{equation}
    \frac{1}{\alpha}\left[ \text{arcsinh}\big(x\big)+x~\sqrt{x^2+1}\right] =t_i.
\end{equation}
which has only one root at $x = 7.47\times10^9$. This root yields the values 
\begin{align}
    \eta_b =&~ \sqrt{\frac{1+x^2}{18\pi\, G\,\rho_\text{max}}} = 2.19\times10^4~\text{s}, \\
    a_b =&~ \left(1+x^2\right)^{-1/2} =7.41\times10^{-9}, \\
    \eta_i =&~ x\,\eta_b =2.96\times10^{12}~\text{s} . 
\end{align} 

\section{Evolution of the cosmological fluid in absence of black holes}\label{app2}

We determine the properties of the cosmological fluid in the absence of black holes, in particular the density $\rho_\text{CF}(\eta)$ and the pressure $ P_\text{CF}(\eta)$. To this end, using the Friedmann equation in terms of the conformal time ($k=\Lambda=0$): 
\begin{equation}\label{friedmann_sola}
\left(\frac{a'(\eta)}{a^2(\eta)}\right)^2 =  \frac{8\pi G}{3c^2} \rho_\text{CF}(\eta), 
\end{equation}
and solving for $\rho_\text{CF}(\eta)$ yields:
\begin{equation}\label{rho_sola}
\rho_\text{CF}(\eta) = \frac{3c^2}{8\pi G}\left(\frac{a'(\eta)}{a^2(\eta)}\right)^2. 
\end{equation}

Taking the time derivative we find
\begin{equation}\label{rho_derivada_sola}
    \rho'_\text{CF}(\eta) =  \frac{3c^2}{4\pi G}\frac{a'(\eta)}{a(\eta)}\left(\frac{a''(\eta)}{a^3(\eta)}-\frac{2a'(\eta)}{a^4(\eta)}\right). 
\end{equation}

From the continuity equation \eqref{casi_sistema_a_resolver2}, we obtain an expression for the pressure $P_\text{CF}(\eta)$:
\begin{equation}
    P_\text{CF}(\eta) = -\frac{1}{3}\frac{a(\eta)}{a'(\eta)} \rho'_\text{CF}(\eta) - \rho_\text{CF}(\eta),
\end{equation}
Since there is no interaction, $Q=0$, and using Eq. \eqref{rho_sola} and Eq. \eqref{rho_derivada_sola}, we get
\begin{equation}
     P_\text{CF}(\eta) =\frac{c^2}{8\pi G} \left(\frac{a'^2(\eta)}{a^4(\eta)} -\frac{2a''(\eta)}{a^3(\eta)}\right). 
\end{equation}

Using the explicit form of the scale factor, the functions result in 

\begin{align}
    \rho_\text{CF}(\eta) &= \frac{3c^2\eta_b^2\,\eta^2}{8\pi Ga_b^2(\eta_b^2+\eta^2)^3},\label{den} \\
    P_\text{CF}(\eta) &=\frac{c^2\eta_b^2\,\eta^2}{8\pi G\,a_b^2(\eta_b^2+\eta^2)^3} - \frac{2c^2\eta_b^4}{8\pi G\,a_b^2(\eta_b^2+\eta^2)^3}\label{presion}. 
\end{align}

As we solved in Sec. \ref{sec5}, the pressure is of the exact same form. This is because the black holes form a pressureless fluid. We can identify two contributions to the pressure $P_\text{CF}$ and using Eq. \eqref{den}, the pressure takes the form
\begin{equation}
    P_\text{CF}(\eta) =\frac{1}{3}\rho_\text{CF}(\eta) - \frac{2\eta_b^2}{3\eta^2}\rho_\text{CF}(\eta).
\end{equation}
As expected, the pressure has a component associated with radiation and a second term that is related to the occurrence of the bounce.

\begin{figure}
    \centering
    \includegraphics[width=\linewidth]{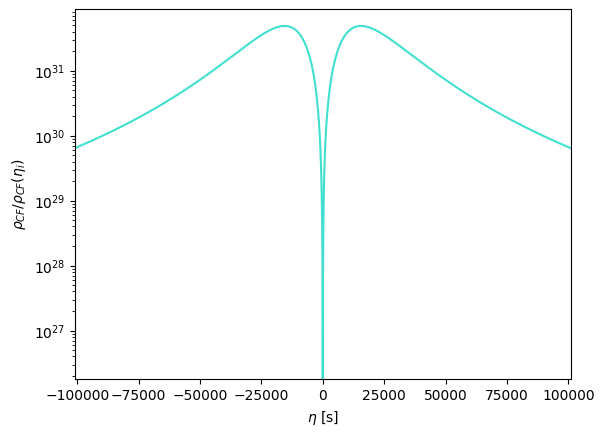}
    \caption{Evolution of the normalized energy density of the cosmological fluid in absence of black holes as a function of the normalized conformal time in the interval $-1.01\times10^5$ s $<\eta<1.01\times10^5$ s.}
    \label{fig:rho-cf-sin-bh}
\end{figure}

Figure~\ref{fig:rho-cf-sin-bh} shows the evolution of the normalized energy density $\rho_\text{CF}(\eta)/\rho_\text{CF}(\eta_\text{i})$, where $\rho_\text{CF}(\eta_\text{i}) = 184.35$ erg\,cm$^{-3}$. For larges values of $\eta$, the energy density becomes null. The energy density presents two maxima at $\eta = \pm 1.55\times10^4$ s and a minimum at $\eta =0$, where it becomes zero. This is a necessary condition for the bounce to happen. Note that the energy density in this case is always positive, except at $\eta=0$ where it annuls. 

\bibliography{Wiley-ASNA}%

\end{document}